\documentclass[a4paper,12pt]{article}
\usepackage{geometry}
\pdfoutput=1

\usepackage[T1]{fontenc} %
\usepackage[swedish,american]{babel}

\usepackage[autostyle]{csquotes}

\usepackage[backend=bibtex,style=authoryear,backref=false]{biblatex} %
\makeatletter

\newrobustcmd*{\parentexttrack}[1]{%
  \begingroup
  \blx@blxinit
  \blx@setsfcodes
  \blx@bibopenparen#1\blx@bibcloseparen
  \endgroup}

\AtEveryCite{%
  }

\makeatother

\let\cite\parencite %
\let\citeN\textcite
\usepackage{hyperref}
\usepackage{color,soul}
\usepackage{adjustbox} %
\usepackage{relsize} %
\usepackage{url}
\def\UrlFont{\small\tt} %
\usepackage{amssymb} %
\usepackage{amsthm}
\usepackage{xfrac}
\theoremstyle{definition}

\usepackage{paralist}
\usepackage{rotating} %
\usepackage{pifont}
\usepackage{wrapfig}
\usepackage{mdframed}
\usepackage{footnote}

\usepackage{afterpage}

\usepackage[shadow,textsize=small]{todonotes}

\usepackage{xspace}
\newcommand{\ie}{\textit{i.e.,}\xspace}
\newcommand{\eg}{\textit{e.g.,}\xspace}

\newcommand{\viceversa}{\textit{vice versa}\xspace}

\newcommand{\CNH}{CN:H\xspace}

\newcommand{\code}[1]{\texttt{#1}}

\begin{document}

\title{Spatiotemporal Modeling of a Pervasive Game {\Large[v5.2]}}
\author{Kim J.L. Nevelsteen\\
Immersive Networking, 
Dept. of Computer and Systems Sciences, \\
Stockholm University
}
\date{}
\maketitle

\newcommand{\DOMAIN}{}
\newcommand{\PROBLEM}{}
\newcommand{\METHOD}{}
\newcommand{\RQ}{}
\newcommand{\SOLUTION}{}
\newcommand{\CONCLUSION}{}
\newcommand{\IMPLICATIONS}{}

\begin{abstract}

Given technology-sustained pervasive games that maintain a virtual spatiotemporal model of the physical world, the implementation must contend with the various representations of space and time. An integrated conceptual model is lacking in the domain of Pervasive Games.
Because Geographical Information Systems and Pervasive Games both make use of the Earth's geography, their problem domains overlap \textit{i.e.,} research found in Geographical Information Systems can be exapted to Pervasive Games. 
To evaluate, the model is applied to the pervasive game, called \mbox{Codename:~Heroes}, as case study. Having an integrated conceptual model, opens up an avenue for the design of the logical and physical model.

\end{abstract}

\section{Introduction}

The focus of this article is on technology-sustained pervasive games~\cite{nevelsteen2015-pervasivemoo} which model physical space and time inside a computer system \textit{i.e.,} virtual; %
where some of the correlations between the physical and the virtual model are maintained. For moving entities, pervasive games can make use of various positioning technologies~\cite{oppermann2009-lbxp} to obtain a spatial reference to the Earth's surface, paired
with a timestamp of the measurement. Elaborate examples of pervasive games, using virtual georeferenced\footnote{When virtual locations use the Earth's surface as a spatial reference, locations can be referred to as `georeferenced'~\cite{adeby2001-principles}.} modeling, are 
Can You See Me Now? and Ambient Wood~\cite{nevelsteen2015-pervasivemoo,dix2005,thompson2003}. %
Given that such games maintain a virtual spatiotemporal model of the physical world, the implementation must contend with the various representations of space and time. Current research pertaining to pervasive games does not provide an integrated spatiotemporal `conceptual'~\cite{abdul2008-3dgis} model for mapping the virtual to the physical and \viceversa.

Because Geographical Information Systems (GIS) and Pervasive Games both make use of the Earth's geography, their problem domains overlap \textit{i.e.,} research found in GIS, on how to model the physical world, can be `exapted'~\cite{johannesson2014-designscience} to Pervasive Games. 
\citeN{dix2005} provide different representations of space and time for pervasive games, but literature with an integrated spatiotemporal conceptual model is lacking.

The approach taken in this article is obtain an integrated spatiotemporal conceptual model through exaptation of \citeauthor{peuquet1994-framework}'s Triad Representational Framework~\cite*{peuquet1994-framework}, combining it with \citeauthor{dix2005}'s three types of space~\cite*{dix2005} and a notion of time. The notions of time by \citeN{langran1992-timegis}, and \citeN{zagal2007}, from the domain of GIS and Pervasive Games, respectively, are equated and incorporated. To evaluate, the model is applied to the pervasive game, called \mbox{Codename:~Heroes}, which maintains a spatiotemporal model of the physical world. Having an integrated conceptual model, opens up an avenue for the design of the logical and physical model~\cite{abdul2008-3dgis}.

\section{Related Work and Problem Statement}

Literature on pervasive games from a technology perspective is scarce~\cite{nevelsteen2015-pervasivemoo}, possibly explaining why no literature was found describing an integrated conceptual model mapping the physical world, through sensors/actuators, to the internal representation of a pervasive game and \viceversa. \citeN{dix2005} do, however, provide different representations of space and time for a pervasive game, %
and so their work is taken into account herein.

Many technology-sustained pervasive games (\eg location-based) maintain a virtual model of physical space and time \eg Can You See Me Now? and Ambient Wood have elaborate models of the physical world~\cite{nevelsteen2015-pervasivemoo,thompson2003}. \citeN{paelke2008-lbgaming} mention the importance of a conceptual model (\eg via GIS) in designing a such a location-based (pervasive) game.
But, the use of GIS in pervasive games is not wide spread; however, cases exist \eg \citeN{bichard2006-backseat} apply GIS objects and spatial data, and imply temporality through the use of prediction. 

The problem is that an integrated spatiotemporal conceptual model is lacking in the domain of Pervasive Games. Such a model should map the physical to the `internal computerized representation' (\ie the virtual), taking into account temporality and that virtual game entities are also objects with relations. In the case where a GIS implementation is combined with a pervasive game (\eg \cite{bichard2006-backseat}), an integrated model can account for redundant copies of the same object. Having a distributed system (\eg \cite{thompson2003}) can exaggerate the redundancy further. 
An initial approach to virtually modeling the physical, might be to map physical space and time directly to virtual space and time. But, such a direct correspondence is an oversimplification of the problem.

\section{Mapping Between the Virtual and Physical}
\label{section:mapping-between-virtual-physical}

A characteristic of a technology-sustained pervasive game is that the game world overlaps with both the virtual and the physical~\cite{nevelsteen2015-pervasivemoo} \ie where the physical needs to be mapped to the virtual and \viceversa.
The approach taken in this article is to exapt \citeauthor{peuquet1994-framework}'s Triad Representational Framework~\cite*{peuquet1994-framework}, from the domain of GIS, allowing for an integrated spatiotemporal model, which includes virtual game entities and their relations.
Then, \citeauthor{dix2005}'s three types of space~\cite*{dix2005} from the domain of Pervasive Games can be combined with a notion of time, tying the virtual to the physical and \viceversa. 

\subsection{Peuquet's Dual Model} %
\label{section:dual}

Considering an overall conceptual representation for geographic phenomena, it can be assumed that any such representation is composed of entities, properties and relationships. 
In a cartographic representation, two dominant views exist~\cite{peuquet1988-dual}: in geometric structure view, the entity is a spatial object, whereas in the graphic image view, the entity is a location. The geometric structure view is referred to as `object-based' (the \textsc{what}) and the graphic image view is referred to as `location-based' (the \textsc{where}). 
Although not entirely distinct, \citeauthor{peuquet1988-dual} presents these views in a unified Dual Model~\cite*{peuquet1988-dual}, on the account that ``neither view is intrinsically better than the other, but are logical duals of each other''. 
Given the \textsc{what} and the \textsc{where}, the Dual Model can be used to form two categories of spatial queries for spatiotemporal analysis: 
\begin{compactitem}
\item \textsc{what}~$\rightarrow$~\textsc{where}~\eg given an object, where is it located?
\item \textsc{where}~$\rightarrow$~\textsc{what} ~\eg given a location, what objects are located there? 
\end{compactitem}

\paragraph{Object-Based Representation}%
According to \citeN{wachowicz1999-ootemporal}, ``any relation defined on a set of entities creates a space~\dots~defining a relation automatically defines a space'' \ie virtual objects of a pervasive games (the \textsc{what}) and the relations between them constitute a space. 
Two spatial relations, that can be used to organize the \textsc{what} of a pervasive game, are a taxonomic hierarchy or a set of objects. The taxonomy allows for the objects to be organized according to their `inheritance'~\cite{peuquet1994-framework} of attributes or thematics \textit{i.e.,} thematic modeling~\cite{abdul2008-3dgis}. %

\paragraph{Location-Based Representation}%
For pervasive games, that make use of positioning technology, a single georeferenced point constitutes a spatial object in the location-based representation of the \textsc{where}~\cite{peuquet1994-framework}, with several georeferenced points forming a bounded areas also adhering to this representation. The spatial relations applicable to pervasive games are metric and topological; spherical distance with the 9-intersection model~\cite{zlatanova2003-topological,abdul2008-3dgis} being sufficient to describe the organization of two objects. 
A directed graph can be constructed in a pervasive game (forming a finite state machine \eg for quests) by having location objects serve as nodes and orientational \code{from-to} relations (directed edges) between nodes~\cite{abdul2008-3dgis}.
\smallskip

By dividing physical space into objects and their locations, the Dual Model allows for the spatial conceptual model of a pervasive game to take into account virtual entities and their relations, including relations to locations. GIS modeling is done as an information model, so the object-based view (\textsc{what}) can be correlated to internal representation of game objects in the engine.

\subsection{Mapping Space to the Physical}
\label{section:mapping_space}

An initial approach to modeling the physical, might be to map the virtual directly to the physical, but such a direct correspondence is an oversimplification of the problem. Sensing and actuating the physical in pervasive games is via instruments, incurring instrumentation error. 
\citeN{dix2005} have captured this error in a model, showing how virtual space can be mapped to and from physical space through `measured space', a representation of space captured in sensors (and actuators\footnote{
Projection is given as an example of mapping directly from virtual space to the physical, but this author would argue that such a projection would also have to cross measured space \eg the projector also requires calibration.
}).
By combining \citeauthor{peuquet1988-dual}'s Dual Model and \citeauthor{dix2005}'s three types of space, virtual objects (spatial object in object-based view) can be related to virtual locations (locations in location-based view) and mapped indirectly (through measured space) to physical locations, and \viceversa. If geometric structure (\eg size, shape, orientation, color and height) is read by sensors, this can be mapped similarly to a virtual object.

\subsection{Peuquet's Triad Representational Framework} %
\label{section:triad}

Since pervasive games can be played in physical or `real-world' time~\cite{zagal2007} (equal to `world time'~\cite[p.34]{langran1992-timegis} in the domain of GIS), change must be accounted for. Temporal objects laid out on a timeline spanning into the future and past make up the `time-based' representation of the \textsc{when}.
\citeauthor{peuquet1994-framework} extends her own Dual Model with the time-based view, forming the Triad Representational Framework~\cite*{peuquet1994-framework} and enabling spatiotemporal queries of the following forms:
\begin{compactitem}
\item \textsc{what}~\texttt{+}~\textsc{where}~$\rightarrow$~\textsc{when} ~\eg given an object at a particular location, when was the last time it changed or appeared? %
\item \textsc{what}~\texttt{+}~\textsc{when}~$\rightarrow$~\textsc{where} ~\eg given an object in a particular time span, what trajectory through space did the object take? %
\item \textsc{where}~\texttt{+}~\textsc{when}~$\rightarrow$~\textsc{what} ~\eg at a particular location, what objects passed by after a particular time? %
\end{compactitem}

\paragraph{Time-Based Representation}%
In a pervasive game, a single unit of `virtual time' \cite{nevelsteenDRAFT-virtualworlddef} is the basic entity %
in the \textsc{when}. According to \citeN{peuquet1994-framework}, all temporal relationships can be divided into three distinct classes:
\begin{inparaenum}[(1)]
\item metrics and topology; 
\item boolean operators; and 
\item generalization.
\end{inparaenum}
Only the first class is needed for pervasive games in this article: temporal distance being ``the length of the interval between any two given locations along a time-line'' and a topology defining how two temporal events relate~\cite{peuquet1994-framework} (also see \citeN{langran1992-timegis} for details).
\smallskip

Both \citeN{peuquet1994-framework} and \citeN{wachowicz1999-ootemporal} make clear that space and time can be viewed from two different perspectives, objective and subjective views corresponding to absolute and relative space-time, respectively. The objective view focuses on space and time geometry as the subject matter, while the subjective view focuses on the objects as the subject matter~\cite{peuquet1994-framework}. From Time Geography it is known that the objective and subjective views are complementary, not contradictory~\cite{peuquet1994-framework}, and that they are integrated~\cite{wachowicz1999-ootemporal}. By treating two representations jointly, certain relationships between objects become more apparent~\cite{wachowicz1999-ootemporal}. Through spatiotemporal analysis, enabled by the Triad, it is possible to go from an objective view to a subjective view. %

\subsection{Mapping Time to the Physical via Measured Time}
\label{section:mapping_time}

If virtual time in the Triad framework is instrumented to and from physical time, then virtual time can be indirectly mapped to physical time (and \viceversa) over `measured time', similar to the technique for physical space in Section~\ref{section:mapping_space}. \mbox{\citeN[p.34]{langran1992-timegis}} notes the difference between physical time and measured time, with respect to GIS literature.

\graphicspath{{img/}}

\section{Evaluating the Integrated Conceptual Model}

In the previous section, \citeauthor{peuquet1994-framework}'s Triad Representational Framework was combined with \citeauthor{dix2005}'s three types of space and a notion of time, forming an integrated spatiotemporal conceptual model that maps the virtual to the physical world and \viceversa. To evaluate the integrated model, the architecture of a pervasive game, called \mbox{Codename:~Heroes (\CNH)}~\cite{nevelsteen2015-pervasivemoo}, is examined as case study.
Contrary to \cite{bichard2006-backseat}, a GIS implementation was not used in CN:H; all representations were modeled and dealt with by the game's implementation. 
A scenario of \CNH is provided here, but with georeferenced zones added to increase the complexity of the example. %

\subsection{Codename:~Heroes Scenario}

Players (or groups of players) roam the physical world on a multi-staged quest, directed by the CN:H game client running on a smartphone with GPS and mobile networking. 
When a player starts their game client for the first time, they are asked to create a profile.
Locations of players in the physical world are obtained through GPS and communicated to the game engine server via mobile networking, along with a timestamp of when the measurement was made. In \CNH, players can encounter physical artifacts with virtual counterparts; if the physical object is interacted with, the virtual counterpart can be accessed simultaneously. Each quest stage leads players from one zone to the next. Each zone is a fictive shape overlaid on the physical world \textit{e.g.,} a point and corresponding radius, or a polygon.
Player movement (\ie their location over time) can be tracked through the visualization of trails of moving game objects. In \CNH, such a visualization was a game master interface prototype that made use of a WebMap.

\subsection{Evaluating Triad}
\label{section:eval_triad}

The Triad Representational Framework is evaluated here first, followed by how it can be mapped to the physical.

\paragraph{Object-Based Representation}%
In \CNH, virtual objects (\eg virtual counterparts of players and artifacts) are organized into a taxonomic hierarchy with inheritance using a \code{parent-child} relation.
One of the objects, in the hierarchy, is specialized as a \code{generic\_admin\_group}, using a \code{member-of} relation to organize child objects into a set. Through inheritance, each child is also a set and represents a unique group, of one or more players, in CN:H. Groups only exist virtually, having no corresponding physical entity, but serve to denote the group's current quest stage.

\begin{figure}
\includegraphics[width=\columnwidth]{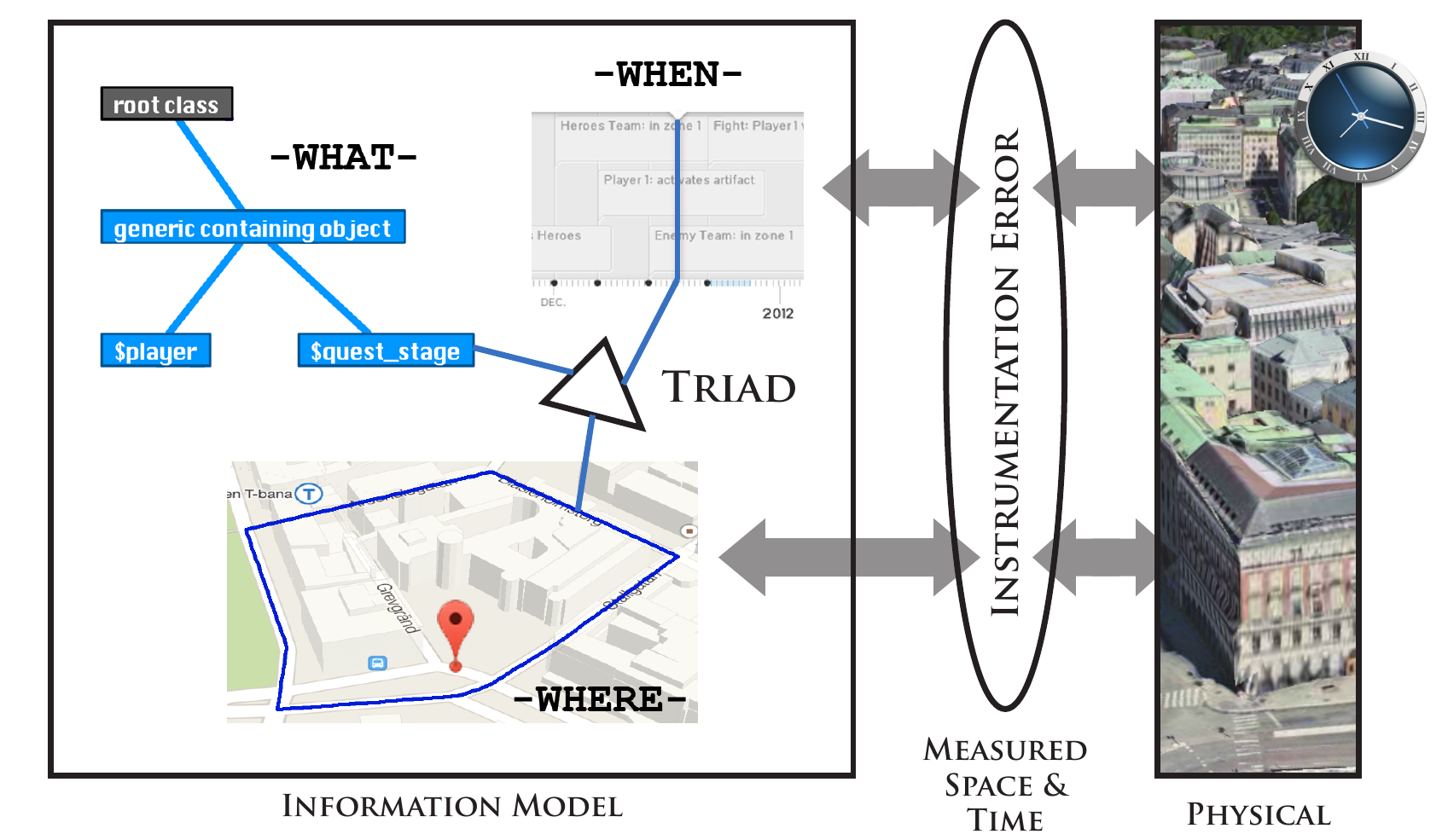}
\caption[]{\label{fig:triad} Triad's \textsc{what}, \textsc{when} and \textsc{where} linked to the physical via measured.}
\end{figure}

Without Triad, the entire object hierarchy would not be a part of the mapping of space and time. Physical location and time would be mapped to virtual location and time, but virtual objects would be left as an implementation detail, possibly leading to redundant copies of the same object  %
\ie a source of inconsistency.
Using Triad, virtual objects are the \textsc{what} having: an identifier attribute, representation specific relations and relations to objects in the \textsc{when} and \textsc{where} (see Fig.~\ref{fig:triad}). Player and artifacts with a corresponding virtual object each store \textsc{when} and \textsc{where} representations in attributes, corresponding to a GPS location and timestamp, respectively. Virtual group objects have a relation to the \textsc{where} corresponding to the group's current quest stage.

\paragraph{Time-Based Representation}%
A `sequent snapshot'~\cite{langran1992-timegis} method was used in \CNH to record GPS measurements \eg a snapshot at regular intervals, recorded on each player's mobile device %
and subsequently communicated to the game engine. In Triad, timestamps are temporal objects in the \textsc{when}, which can be used to create the timeline depicted on Fig.~\ref{fig:triad}.

\paragraph{Location-Based Representation}%
Two types of location objects are present in \CNH: a single georeferenced coordinate and an area (`zone') consisting of a georeferenced boundary (depicted as a red flag marker and blue hexagon on Fig.~\ref{fig:triad}, respectively). GPS describing the World Geodetic System is the basis for each georeference, with latitude and longitude~\cite{paelke2008-lbgaming} dividing up the Earth's space into a discrete field of cells~\cite{adeby2001-principles}. Altitude is often discounted when using GPS, implying it extends indefinitely in that direction. If altitude is taken into account, it is often sufficient to consider altitude as `elevational layers'; a so-called 2\sfrac{1}{2}D representation, rather than describing a volumetric 3D representation~\cite{adeby2001-principles}. In CN:H, 2\sfrac{1}{2}D was sufficient for player and artifacts, and zones were left as 2D with implied infinite altitude. Each quest stage was a zone corresponding to a bounded area in the physical world. 
Using a directed graph between zones, different quest stages were interconnected to form a multi-staged quest.
Location objects the in \textsc{where} with updates can either overwrite previous values or form of a log of values; in \CNH it was the latter.
\smallskip

\newcommand{\Question}[1]{\textit{\small #1}}

Because Triad provides an integrated spatiotemporal conceptual model for \CNH, questions relevant to the game scenario can be translated into corresponding spatiotemporal queries. The questions \Question{``Show me the movement trail of \mbox{Player-1}?''} and \Question{``What is the distance between one game object and another entity or zone?''} translate into a query of the form: \textsc{what}~\texttt{+}~\textsc{when}~$\rightarrow$~\textsc{where} \ie a range query and a distance calculation of two subsequent queries, respectively.
Questions \Question{``What players were in \mbox{Zone-A} at 7pm?''} and \Question{``Who was playing in the area of Stockholm on Saturday?''} translate into form: \textsc{when}~\texttt{+}~\textsc{where}~$\rightarrow$~\textsc{what}. And, lastly, \Question{``What time did \mbox{Player-1} enter \mbox{Zone-A}?''} results in a temporal query of the form: \textsc{what}~\texttt{+}~\textsc{where}~$\rightarrow$~\textsc{when}. 
For \CNH, results of the spatiotemporal queries were visualized as space-time paths, in combination with a timeline~\cite{guerrero2014-gm}, on a WebMap game master interface~\cite{nevelsteen2015-pervasivemoo}. Other visualizations are also possible \eg a space-time cube~\cite{langran1992-timegis}.

\subsection{Evaluating the Mapping to Physical and \viceversa}
\label{section:eval_mapping}

In this section, \citeauthor{dix2005}'s three types of space in combination with Triad is evaluated. 
In \CNH, it was chosen to have the game engine hold the primary copy of each object, with each game client holding a redundant copy %
\ie updates were sent directly to the game engine server and disseminated to game clients. 
Because of Triad, redundant copies of objects on various devices can be accounted for and still fall in the object-based representation (the \textsc{what}). In the game engine, virtual objects are related to the \textsc{where} and \textsc{when}.

After creating a profile, a player's location can already be read from their device;
sensors pick up GPS signals, measuring longitude and latitude~\cite{nco2013-gps}.
The location  (the \textsc{where}) in physical space measured by the sensors, constitutes measured space. %
Measurements saved to device storage (with possible error at this stage or subsequent stages of storage, due to limitations of internal representation, transmission, rounding or truncation), constitute virtual space.

Although it is possible obtain a snapshot of time (the \textsc{when}) from the GPS signal~\cite{nco2013-gps}, in \CNH, a timestamp was formed using device time, which was (by default) synchronized with Coordinated Universal Time. Device times constitute measured time. 
Time snapshots (possibly different for each device) are transmitted to the server and reconciled with the game engine time, constituting virtual time. %

In \CNH, it was the WebMap game master interface ~\cite{nevelsteen2015-pervasivemoo} that was responsible for visualizing player positions in the game world. The WebMap connected simultaneously with both the game engine and OpenStreetMap~\cite{osmf2004-map}; the game engine to obtain a subjective view of the model through spatiotemporal queries and the OpenStreetMap for map data. Virtual space and time are transferred from the game engine into the WebMap display. For each player, a line of dots is drawn on the display representing the player's movement trace, with each dot (the \textsc{where}) marked with the player's ID (the \textsc{what}) and a timestamp (the \textsc{when}). Since the WebMap can not be physically misaligned (\eg like sensors or a robotic arm), this author would argue that the person reading the display and checking the physical space is the instrument (possibly misaligned with the physical world), constituting measured space and time \eg not finding the player where expected. And, of course, it is not possible that measured space and time are exact when compared to the infinite resolution of physical space and time.

\subsection{Solution Maturity of GIS and Pervasive Games}

The domain of GIS has many solutions that can be exapted to Pervasive Games. The reason for this is that the solution maturity of GIS is high and that of Pervasive Games low, with respect to the Earth's geography. There are still some solutions that can be exapted to GIS from Pervasive Games \eg with respect to real-time updates.

\CNH only makes uses of discernible boundaries, when indiscernible boundaries are more difficult~\cite{abdul2008-3dgis} \eg enclosing an area of a gaseous element. And, in Pervasive Games, the usage of indiscernible boundaries has already been reported \eg by using a fuzziness threshold when modeling team positions~\cite{thompson2003}. Also in \CNH, the temporal aspects used were minimal; many of the complex spatiotemporal changes possible in the physical world were not modeled in the game \eg changes in landscapes or building architecture. Because \CNH was staged over a period of weeks, not months or years, such changes were not encountered. There is much complexity with regards to temporality handled by GIS (see \cite{langran1992-timegis}), so as pervasive games become more advanced, similar complexity might need to be addressed. In \CNH, data was recorded in an objective view of space and time. Recording data in a subjective view should have been considered as an optimization~\cite{peuquet1994-framework,langran1992-timegis}. \citeN{worboys2004-gisbook} discuss a brief history of time, specifying ``object lifetimes'' and ``events, actions and processes'' as later subsequent evolutions of the snapshot method used in \CNH.

\section{Conclusion}

This article provides an integrated spatiotemporal conceptual model, through the exaptation of \citeauthor{peuquet1994-framework}'s Triad Representational Framework to the domain of Pervasive Games, combining it with \citeauthor{dix2005}'s three types of space and a notion of time. The notions of time by \citeauthor{langran1992-timegis}, and \citeauthor{zagal2007}, from the domain of GIS and Pervasive Games, respectively, are equated and incorporated. The model is integrated, taking into account temporality (Section~\ref{section:triad}) and enabling spatiotemporal queries (Section~\ref{section:eval_triad}). The model takes takes into account virtual objects and their relations (Sections~\ref{section:dual} and \ref{section:eval_triad}) and accounts for redundant copies of the same object (Section~\ref{section:eval_mapping}). Rather than attempt to map the virtual directly to the physical, \citeauthor{dix2005}'s three types of space is used to handle measured space and time (Sections~\ref{section:mapping_space}, \ref{section:mapping_time} \& \ref{section:eval_mapping}). Having an integrated conceptual model, opens up an avenue for the design of the logical and physical model~\cite{abdul2008-3dgis}.
The solution maturity of GIS is high and that of Pervasive Games low, with respect to the Earth's geography; depending on the functionality required, there are perhaps still many concepts from the domain of GIS that can be exapted to Pervasive Games.

\section*{\textsc{Acknowledgements}}
Thank you to Martijn Meijers, Annika Waern, Barry Brown and Inger Ekman for their critical reviews. 
Research was made possible by the Department of Computer and Systems Sciences and a grant from the Swedish Governmental Agency for Innovation Systems to the Mobile Life Vinn Excellence Center.

\renewcommand*{\bibfont}{\small}
\renewcommand*{\UrlFont}{\ttfamily\smaller\relax}

\printbibliography[title=\textsc{References}]

\end{document}